# Numerical Investigation of Supersonic Turbulent Flow Over Open Cavities

Roshan Sah[*], Sanchari Ghosh

Department of Aerospace Engineering, Indian Institute Of Technology Kharagpur, Kharagpur, India

**Email address:**
sahroshan11@gmail.com (R. Sah), sanchari.ghosh732@gmail.com (S. Ghosh)
[*]Corresponding author

**Abstract:** The study of cavity flow is one of the most important research topics of unsteady aerodynamics. Supersonic turbulent flows over a cavity are mostly encountered in missiles, turbomachinery, and high-speed aircraft. The turbulence inside the cavity gives rise to excess drag, acoustic waves, pressure fluctuations, and vibrations which may lead to numerous problems like excess fuel consumption, failure of missile trajectory or mechanical parts, and aerodynamic heating. We conduct numerical simulations to investigate the flow and compare our results to existing experimental data to show quantitative validation. We then investigate the effect of the Mach number of the turbulent supersonic flow on the pressure contours and the vortical structures inside the cavity and the subsequent effect on other flow parameters like acoustic waves. Thereafter, we propose two modifications of the cavity geometry, a) slanted edges and b) smoothened corners with slanted edges, to improve the aerodynamic performance. ANSYS ICEM tool has been used for the fine mesh generation of our cavity geometry and all the simulation was run on ANSYS fluent software. The K-ω SST turbulence model was used for the simulation as it can capture the near-wall property with reasonable accuracy and a grid independence study was carried out to find the correct solution of the Navier-Stokes equation was proceeding by the solver. The parameters like pressure contours, streamline pattern, coefficient of pressure distribution and sound pressure level (SPL) has been found and compared for both modified and unmodified open cavity at different Mach number. The modifications suggested show a significant improvement over the open cavity designs in the mean pressure distribution and sound pressure level distributions.

**Keywords:** Cavity, Vorticity, Aspect ratio, Aerodynamic Characteristics Pressure Distribution, Mach Number, Supersonic Nomenclature

## 1. Introduction

High speed flows inside cavities have numerous applications in aerospace and aeronautical field. Especially in the aerospace field, turbulent, unsteady and complex flow phenomena become predominant part of the process. In aeronautical application, interior carriage stores like landing gear, release carriage stores like weapons and bombs, all are cavity configuration and in a scramjet engine, cavity flame holder is used for proper ignition of flame by reducing the ignition delay time over a range of operating conditions.

At high speed fluid flow over the cavities, an unsteady disturbance and complex flow field are generated around the cavity. These disturbances and flow fields inside the cavity cause fluctuations in the static pressure distribution which result in development of large pressure gradients and relatively high sound pressure levels along the cavity. Problems like shear layer instabilities, vortical instabilities, turbulent boundary layer separation, shock-wave/boundary-layer interaction are produced when supersonic flow occurs over the cavity, which can affect the health of the aircraft and the stores.

Based on flow field structure inside the rectangular cavity, Stallings *et al.* [1] divided a supersonic cavity flow into three configurations depending on ratio of the aspect ratio (length to depth *i.e.* L/D). In an open cavity, L/D is less than 10, whereas L/D between 10 and 13, is referred to as a transitional cavity configuration and the flow in a closed cavity corresponds to L/D greater than 13 [1].

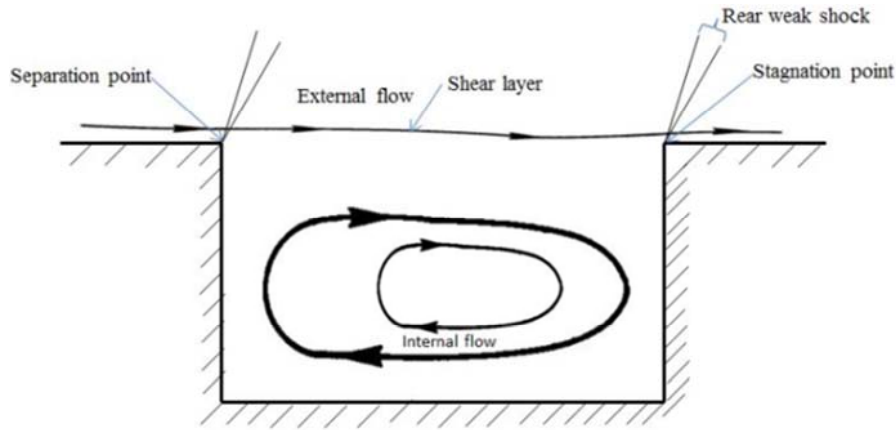

*Figure 1. Supersonic cavities flow over an open cavity.*

Schematic diagram of open cavity configuration for supersonic flow is shown in figure 1. The distribution of static pressure fluctuation inside the rectangular open cavity was experimentally determined by Tracy & Plentovich [2] and Plentovich et al [3]. Investigations were also carried out by researchers Yang *et. al.* [4] in 2008 on supersonic and transonic flows over open and closed cavities. In-detail investigation, both experimental and numerical, was carried out to find the effect of Mach number and aspect ratio of the cavity on the streamlines as well as the pressure distribution inside the cavity. The highest Mach number investigated was 1.5 by Yang *et. al.* [4]; subsequently in Sridhar et al. [5], investigated the aerodynamics of an open cavity with L/D ratio 8 when placed in a supersonic flow of Mach number 2.

Experimental work on cavity flow of Rossiter [6-8] and Rossiter & Kurn [9] was followed up by many researches for instance: Bueno et al [10], who proved that $C_p$ increases with increase in cavity length as well as the effect of different L/D ratios on supersonic flow field. The aspect ratio L/D was found to be the main factor affecting the acoustic oscillations although the ratio did not seem to affect the shock impingement effect.

In the numerical simulations, many models have been used over the years to replicate the physics of the flow field. While some of them produced satisfactory resemblance, the rest had considerable variation compared to the experimental data. The URANS model was used to numerically investigate both 2 D and 3 D flows by Shieh & Morris [11]. Shieh & co-workers also subsequently used the well-known k-ε model to understand flow field at a Mach number of 1.5 [11, 12]. They found the mass transfer between cavity and free flow is the primary cause of the flow instabilities. Large Eddy Simulations (LES) was used to simulate supersonic flows by Rizetta et al. [13] whereas Detached Eddy Simulation (DES) was used by Hamed et al. [14] and they found that the vortical structures are initially formed near the leading edge of the cavity which subsequently get pushed towards the trailing edge by the flow.

In this article, we examine the effect of transonic and supersonic turbulent flow over an open cavity using three dimensional numerical simulations using ANSYS FLUENT. In addition, we propose two modifications, 1) slanted edges and 2) smoothened corners with slanted edges which improve the mean pressure distribution as well the sound pressure level distribution over the cavity.

## 2. Numerical Simulation

### *2.1. Model Geometry*

We have chosen a cavity with aspect ratio 6, (L=0.06 m, D=0.01 m, W=0.055 m). The inlet is situated 15D ahead of the leading edge of the cavity and the outlet is at a distance of 30D from the trailing edge of the cavity and the domain height is 30D from the cavity lip. The geometry of the cavities considered in our analysis is shown in figures 2 (a-d).

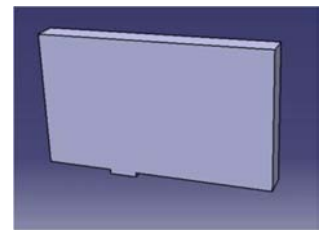

(a) Open rectangular cavity (L/D=6) model with flow domain.

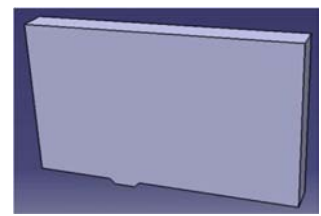

(b) Modified corner of rectangular open cavity (L/D=6) model with flow domain.

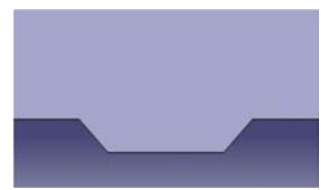

(c) Zoomed view of modified rectangular corner cavity L/D=6.

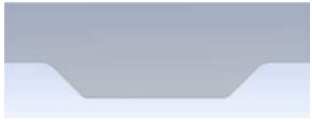

(d) Zoomed view of smoothened modified cavity L/D=6.

*Figure 2. Design model: (a) open cavity (L/D=6), (b) Modified open cavity, (c) Zoomed view of modified open cavity. (d) Zoomed view of smoothened modified cavity.*

Subsequently, keeping all the other boundary conditions the same, we modified the cavity geometry by inclining the leading and trailing edge walls at an angle of 45 degrees to the cavity bed as shown in figure 2 (b) & 2 (c). After that we have introduced curvatures 2 (d) at the leading and trailing edges of the cavity (5 mm in radius) and also at the bottom edges of the cavity (3 mm in radius) to analyze the effect of smoothening.

### 2.2. Meshing

For the open cavity shown in figure 3 (a), a structured mesh was created in which we have refined the mesh inside and adjoining areas of the cavity and comparatively a coarse mesh in the rest of the flow domain. The fine mesh region has been discretized with maximum grid spacing of 1 mm to capture the wall effect and the boundary layer on the flow field. And the rest of the flow domain has been discretized with maximum grid spacing of 2 mm which gives satisfactory results in shock capture.

The zoomed view of a fine mesh rectangular cavity is shown in figure 3 (b) which showed us a fine mesh pattern with equal grid spacing.

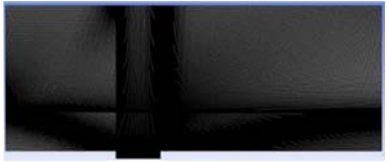

(a) Fine meshing of rectangular open cavity model with flow domain.

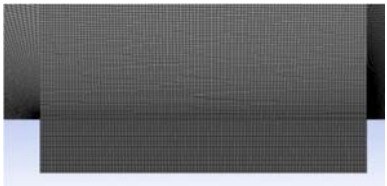

(b) Zoomed view of mesh of rectangular open cavity.

*Figure 3. Meshing: (a) Fine mesh of rectangular open cavity, (b) Zoomed view of the mesh.*

The fine mesh region has been discretized with maximum grid spacing of 0.5mm to capture the wall effect and the boundary layer on the flow field. And the rest of the flow domain has been discretized with maximum grid spacing of 1 mm which gives satisfactory results in shock capture.

Similarly for the modified geometries 4 (b) and 4 (c) we have opted for finer meshing near and inside the cavity in comparison to the rest of the flow domain. Additionally, here in the finer mesh we have taken maximum grid spacing as 0.5 mm to better capture the effects of the inclined walls.

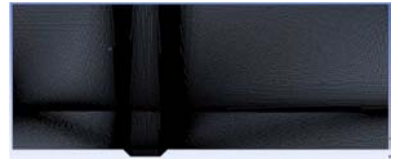

*(a)* Fine meshing of modified corner of rectangular open cavity model with flow domain.

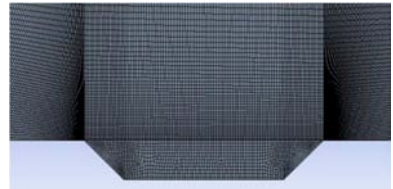

*(b)* Zoomed view of mesh of modified corner of rectangular open cavity.

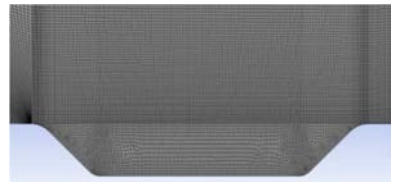

*(c)* Zoomed view of mesh of smoothened open cavity.

*Figure 4. Meshing: (a) Fine mesh of modified corner rectangular open cavity, (b) Close up view of the mesh of modified cavity & (c) Close up view of mesh of smoothened open cavity.*

### 2.3. Computational Methodology

The 2-D Reynolds-Averaged compressible NSE (Eq. 1) is solved by utilizing a finite volume spatial discretization (FVSD).

$$\frac{\partial M}{\partial \xi}+\frac{\partial N}{\partial \eta} = \frac{1}{Re}\left(\frac{\partial M_v}{\partial \xi}+\frac{\partial N_v}{\partial \eta}\right) \qquad (1)$$

Where M and N are convective fluxes and $M_v$ & $N_v$ are diffusive fluxes. The computations were performed with second order accurate scheme in space and the convective fluxes are modeled using $2^{nd}$ order implicit Roe-FDS model [15, 16] which has been reported to capture experimental results more closely. The two dimensional implicit steady density based solver has been in this simulation to capture the compressible effect of the flow.

### 2.4. Turbulent Model and Boundary Condition

The K-ω model [16] was used for the modeling of the turbulence in the cavity flow where two additional transport equations, turbulent kinetic energy K and the rate of dissipation of turbulence kinetic energy per unit kinetic energy ω are solved. The shear-stress transport (SST) form is solved as shown in equations (2-6). The advantage of the SST model is that the high Reynolds number region as well as near-wall regions can be modeled with reasonable accuracy. The formulation of the model SST used herein is:

$$\frac{\partial(\rho k)}{\partial t} + \frac{\partial(\rho U_j k)}{\partial x_j} = P_k - \beta^* \rho k\omega + \frac{\partial}{\partial x_j}\left(\Gamma_k \frac{\partial k}{\partial x_j}\right) \quad (2)$$

$$\frac{\partial(\rho\omega)}{\partial t} + \frac{\partial(\rho U_j \omega)}{\partial x_j} = \frac{\gamma}{\nu_t} P_k - \beta\rho\omega^2 + \frac{\partial}{\partial x_j}\left(\Gamma_\omega \frac{\partial \omega}{\partial x_j}\right) + 2\rho\sigma_{\omega 2}\frac{1}{\omega}\frac{\partial(k)}{\partial x_j}\frac{\partial(\omega)}{\partial x_j} \quad (3)$$

$$\Gamma_k = \mu + \mu_t/\sigma_k, \Gamma_\omega = \mu + \mu_t/\sigma_\omega \quad (4)$$

$$P_k = \tau_{ij}\,\partial U_i/\partial x_i \text{ and } P_k = \min(P_k, C_{1\epsilon}) \quad (5)$$

The coefficients $\varphi_1$ and $\varphi_2$ of the model are functions of:

$$\varphi = f1\varphi_1 + (1-f1)\varphi_2 \quad (6)$$

Where the coefficients of the model k-ω are shown in table 1.

***Table 1.** k-ω model coefficient values.*

| | | |
|---|---|---|
| $\sigma_{k1}$=2.0 | $\kappa$=0.41 | $\gamma_1$=0.5532 |
| $\sigma_{\omega 1}$=2.0 | $C_1$=10 | $\gamma_2$=0.4403 |
| $\beta^*$=0.09 | $\beta_1$=0.075 | $\beta_2$=0.0828 |
| $\sigma_{k2}$=2. | $\sigma_{\omega 2}$=1.168 | |

The Sutherland viscosity law is employed in the computation. The no-slip condition was imposed at the walls (eq. 7) where n denotes wall-normal direction. The wall temperature was assumed to be isothermal at 288 K, and outflow condition was specified at the exit of the domain.

$$u=v=0,\ \partial P/\partial n=0 \text{ and } T_1 = T_{aw} \quad (7)$$

***Table 2.** Boundary conditions and its values.*

| Parameter | Values |
|---|---|
| Operating pressure | 101325 pa |
| Free stream stagnation temperature | 288K |
| Solver type | Density based |
| Turbulence model | SST K-ω model |
| Operating Mach Numbers | 0.6, 1.2, 1.5, 2, 2.5 |
| Operating fluid | Ideal Gas |
| Wall condition | No slip wall |

Sound pressure level (SPL) values are calculated from the following equation (eq. 8, 9 & 10). An average pressure coefficient (Cp) values are also presented to compare the result between the rectangular cavity and the modified corner in the rectangular cavity. Fast Fourier Transform (FFT) is used to find oscillation frequencies numerically. Velocity and pressure contours help to understand cavity physics.

$$SPL = 10\log_{10}[P^2/q^2] \quad (8)$$

$$\bar{P}^2 = \frac{1}{tf-ti}\int_{ti}^{tf}(P-\bar{P})^2 dt \quad (9)$$

$$P = \frac{1}{tf-ti}\int_{ti}^{tf}(P-\bar{P})dt \quad (10)$$

Where P is the pressure value at each point, q is reference sound pressure level value which is equal to $2\times 10^{-5}$ Pa and $t_f - t_i$ = time spacing.

## 3. Grid Independence Study

An important part of any numerical investigation and the subsequent results is the independence of the results on the grid resolution. This proves the correct solution of the Navier Stokes equations is being carried out by the solver. If the parameters vary rigorously with the type of grid deployed it proves that the solution has intrinsic error and has to be further modified so that the results emulate the physics correctly. Hence a grid independence study has been carried out with the following mesh types.

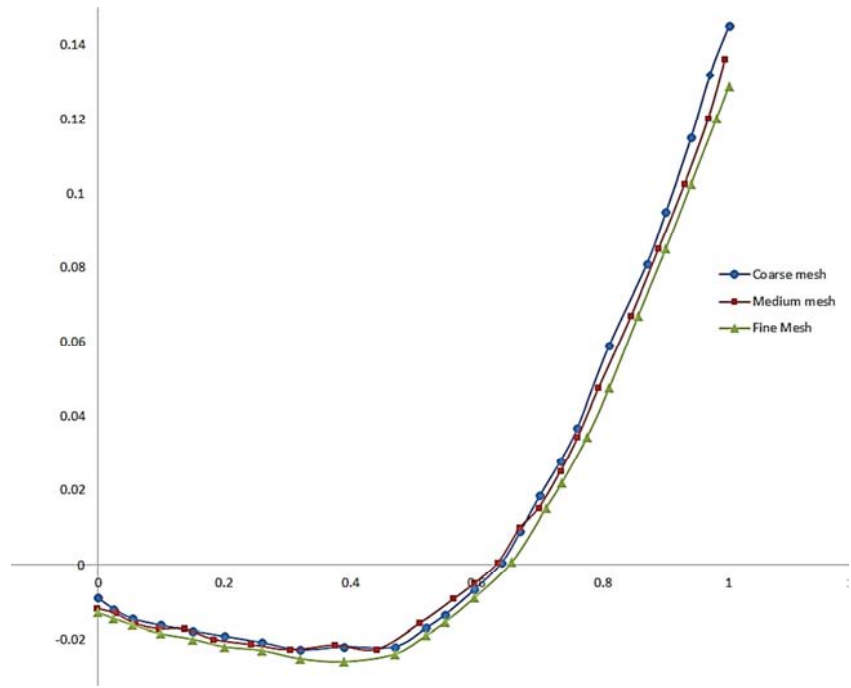

***Figure 5.** Grid independence study: Comparison of pressure distribution with varying mesh type.*

*Table 3. Number of nodes and elements of each mesh type.*

| Mesh Type | Number of nodes | Number of elements |
|---|---|---|
| Coarse mesh | 70001 | 69375 |
| Medium mesh | 115801 | 115000 |
| Fine mesh | 145926 | 145000 |

On plotting the $C_p$ distribution (Figure 5) inside the cavity for the different mesh types deployed we see that the plots align diligently almost everywhere inside the cavity, thus proving the parameter is invariant with respect to grid resolution [17-19].

Thus, as the solutions are grid independent, we proceed to validate the numerical data with existing experimental and numerical results. For doing so we use the fine mesh type with the given number of nodes and elements.

## 4. Validation

To validate our numerical results we tallied the coefficient of pressure distribution inside the cavity (both rectangular and modified of L/D=6) as calculated with pre-existing experimental and numerical data on a similar cavity under the same boundary conditions in the work done by Dang Guo Yang et al [4]. The resultant graphical representation shows similar trends of coefficient of pressure distribution over the rectangular open cavity. There is a slight increment in the values of $C_p$ for the computational case than that of experimental data and the difference can be attributed to the slight incorrectness in the estimation of mixing length in the k-ω turbulence model. The pressure contour of modified and unmodified open cavity with streamline is shown in figure 6 (a) & (b), whereas in 6 (c) show the comparison of $C_p$ distribution of modified and unmodified open cavity with previously done experimental data. After validating numerical procedures for tallied graphs, the same procedure can be used for the simulation of supersonic flow over cavity of modified and unmodified configuration.

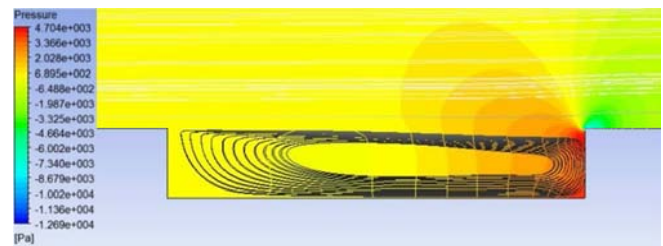

(a) Open cavity.

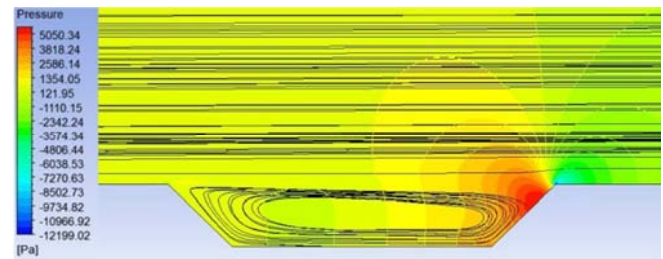

(b) Modified open cavity.

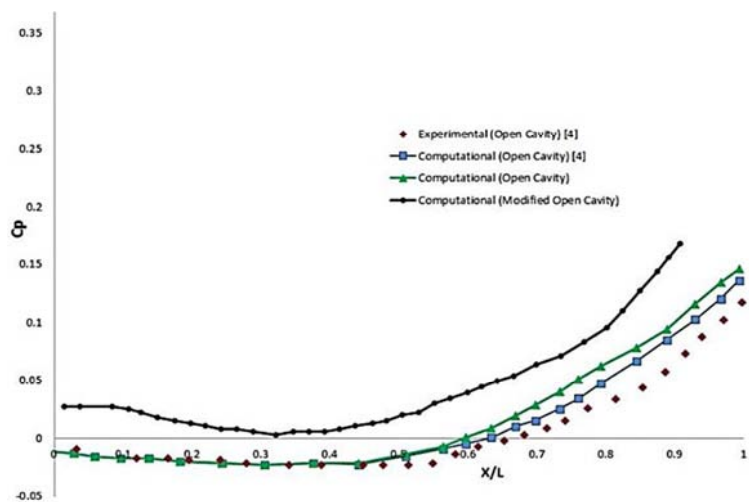

(c) $C_p$ distribution inside cavity.

***Figure 6.*** *Pressure distribution: (a) Open cavity, (b) Modified open cavity, (c) Comparison of Cp distribution.*

## 5. Result And Discussion

### 5.1. Open Cavity

Figure 7 shows the flow over the rectangular open cavity with L/D=6 at supersonic speeds. At supersonic speeds, a shear layer is generated over the cavity and then we observe the occurrence of expansion waves at the front edge of the cavity. In the central region of the cavity, an oblique shock wave is found and finally as the flow exits the trailing edge of the cavity a second shock wave is observed. As the Mach number is increased, strength of shock wave also increases at both leading and trailing faces of the cavity, resulting in the development of a large pressure gradient across the cavity [17, 20].

The high pressure region behind the first shock interacts with the low pressure region near the trailing edge setting up a central vortex in the cavity flow field. This central vortex formation has been shown by the stream line pattern at each pressure contour figure. And at each Mach number, positive pressure distribution is found.

From figure 7, we can observe that, as the Mach no. of the flow is increased, the central vortex formed gets stretched toward the trailing face of the cavity. Thus, shifting in vortex, results in minimum pressure location ahead of the trailing face of the cavity. And this minimum pressure is positive at each Mach number, which result in a positive coefficient of pressure ($C_p$) at each location of the cavity.

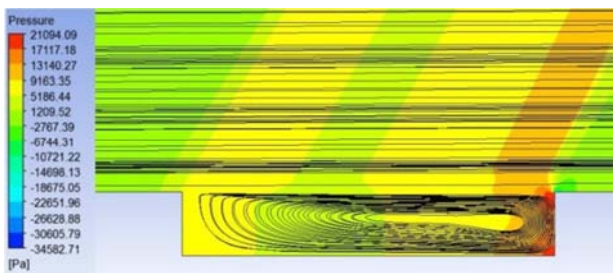

(a) Streamline pattern with pressure contours at M=1.2.

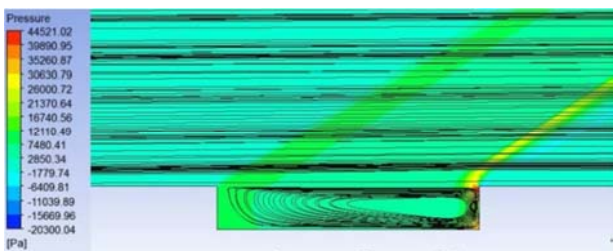

(b) Streamline pattern with pressure contours at M=2.

***Figure 7.*** *Pressure Contour on open cavity at different Mach number with streamline.*

### 5.2. Modified Open Cavity

Figure 8 shows the flow over the modified rectangular open cavity of L/D=6 at supersonic speeds. In a modified open cavity, both expansion and oblique shock waves are formed at the leading and trailing edge of the cavity similar to the open cavity.

From the pressure contours, we can see that there is a high pressure zone at the rear edge of the cavity and a low pressure zone near the front edge of the cavity because of the shock wave and the expansion fan respectively. In the modified open cavity, a more centered vortex is formed that can be represented by a streamline pattern and it's not localizing towards the trailing face. This central vortex results in minimum pressure location at middle flow of the cavity. And this minimum pressure can result in positive or negative pressure as Mach number changes, which result positive or negative coefficient of pressure ($C_p$) at different location of the cavity. Negative pressure coefficient distribution across the cavity can result in decrease in drag [21].

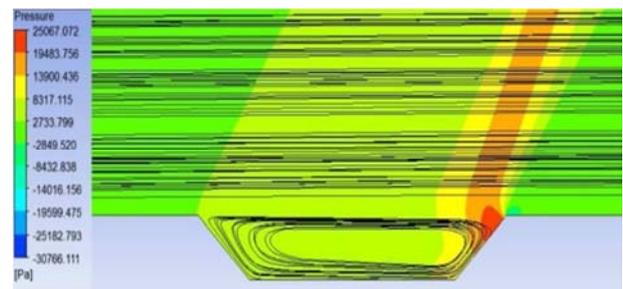

(a) Streamline pattern with pressure contours at M=1.2.

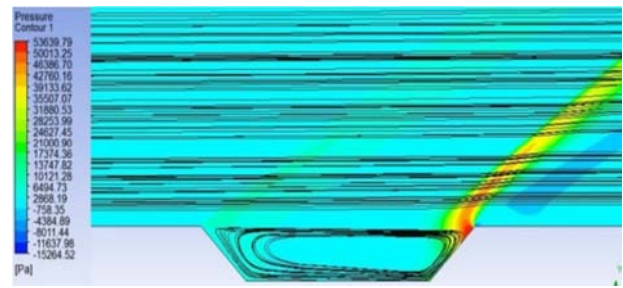

(b) Streamline pattern with pressure contours at M=2.

***Figure 8.*** *Pressure Contour on modified open cavity at different Mach number with streamline.*

### 5.3. Smoothened Cavity

Figure 9 shows the pressure contours and the streamlines inside the cavity on smoothening the edges at the mach numbers 1.2 and 2.

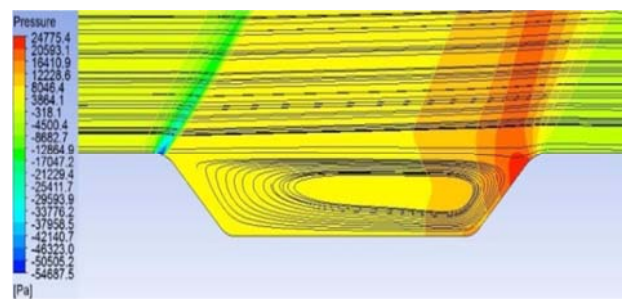

(a) Streamline pattern with pressure contour at M=1.2.

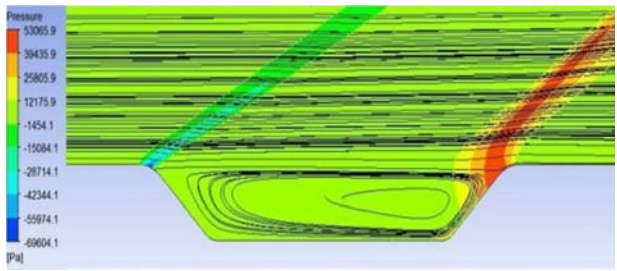

(b) Streamline pattern with pressure contour at M=2.

*Figure 9. Pressure Contour on smoothened open cavity at different Mach number with streamline.*

We see that after smoothening of the edges there has been a further decrement in the pressure inside the cavity and this decrement has increased with the increase in Mach number. On comparison of the pressure contours and the streamlines with our initial cavity we see that the location of the shock at the trailing face of the cavity has travelled upstream from the lip of the cavity and the shock is created a little below the lip of the cavity. Also the inclination of the shock increases with the Mach number thus the shock strength has decreased even though the Mach number of the flow has increased. The vortex inside the cavity is similar to that of the cavity with sharp slant sides, centered and upstretched.

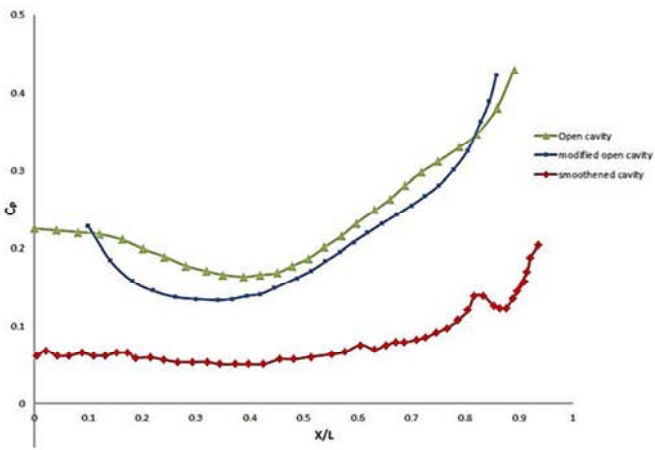

(a) $C_p$ distribution at M=1.2.

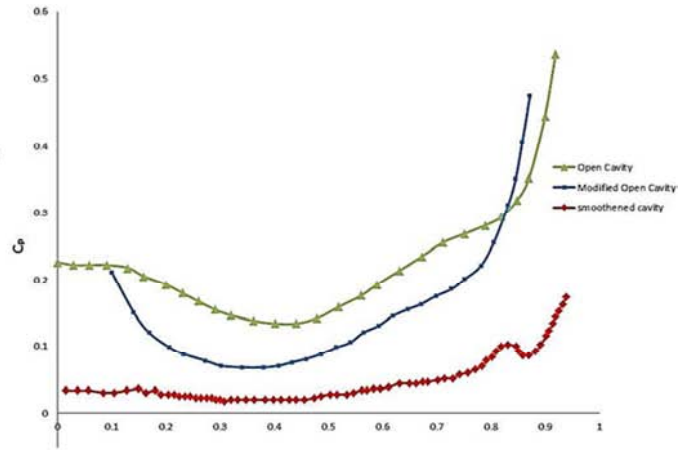

(b) $C_p$ distribution at M=1.5.

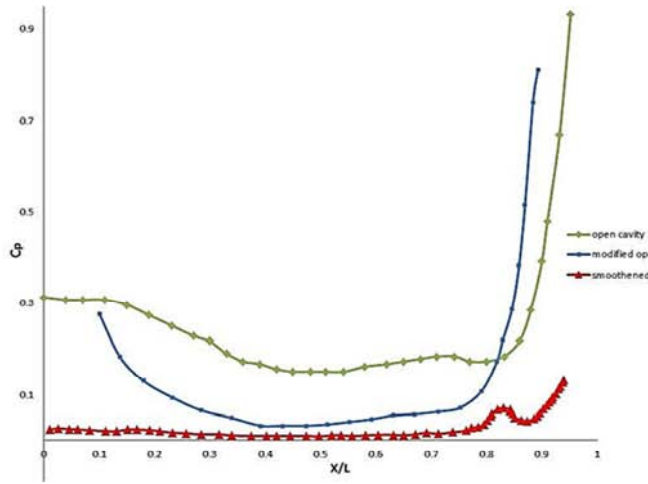

(c) $C_p$ distribution at M=2.

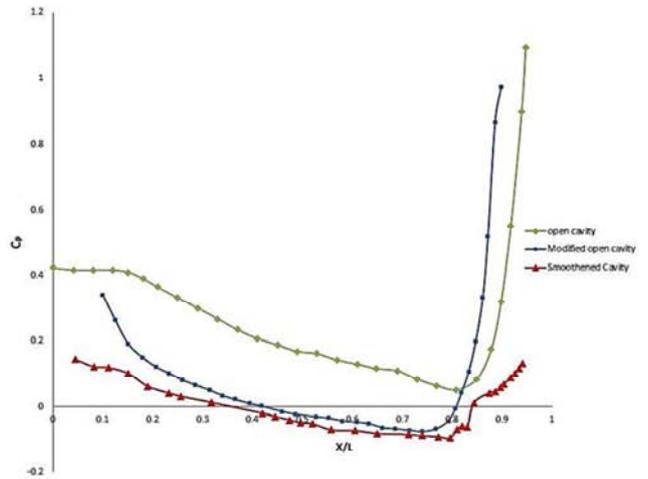

(d) $C_p$ distribution at M=2.5.

*Figure 10. Comparison of pressure coefficient of open cavity and Modified open cavity at different Mach number.*

### 5.4. Comparison of $C_p$ Distribution of Open and Modified Open Cavity

Figure 10 shows the graphical representation along the length of the cavity. The green line represents the unmodified cavity $C_P$ distribution, the blue one represents the modified one and the red one represents that of the smoothened one. The first thing that catches the eye is that in both the cases of the rectangular and modified cavity the graphs have a similar trend in case of all the Mach numbers. But for the smoothened cavity the $C_P$ has a more linear variation along the cavity. Also, the value of the property decreases overall due to the applied modification. Like the unmodified cavity, for the sharp edged modified cavity, the least $C_P$ zone lies at the central region of the cavity due to the presence of vortex in both the cases as shown in Figures 7 and 8. But, in the latter, we see, the $C_P$ decreases at the central part of the cavity steadily with the increment of Mach number. In fact at Mach number 2.5, the property attains a negative value. As for example, for Mach number 1.2, at the point x/L=0.3 inside the cavity, the value of $C_P$ is between 0.1 and 0.15. This value at this particular position has continuously decreased as we have increased the Mach number and has gotten a negative value at Mach number 2.5. The decrease in $C_P$ ensures an overall decrease in drag, which in turn increases with increase in Mach number. Also negative $C_P$ denotes the onset of opposite circulation direction.

After smoothening the corners, we see a further decrement in the value of pressure coefficient all along the cavity. In fact, unlike the one with the cavities with sharp edges, no sudden drop is seen near the central portion of the cavity. There is just a linear variation of a decreased value of the pressure coefficient for all the Mach numbers. This indicates further drag decrement inside the cavity.

*5.5. Comparison of SPL Distribution Across Cavity:*

Figure 11 represents the SPL distribution across the cavity length for different Mach numbers. The blue line represents the property value for the unmodified cavity and the red one represents that for a modified one. The green one is the graphical representation of SPL for the smoothened cavity.

For any open cavity, there is a pressure difference in existence due to high pressure at the trailing face and low pressure at the leading face. That difference leads to the onset of pressure oscillation inside the cavity which in turn is a measure of acoustic noise which is measured by the SPL.

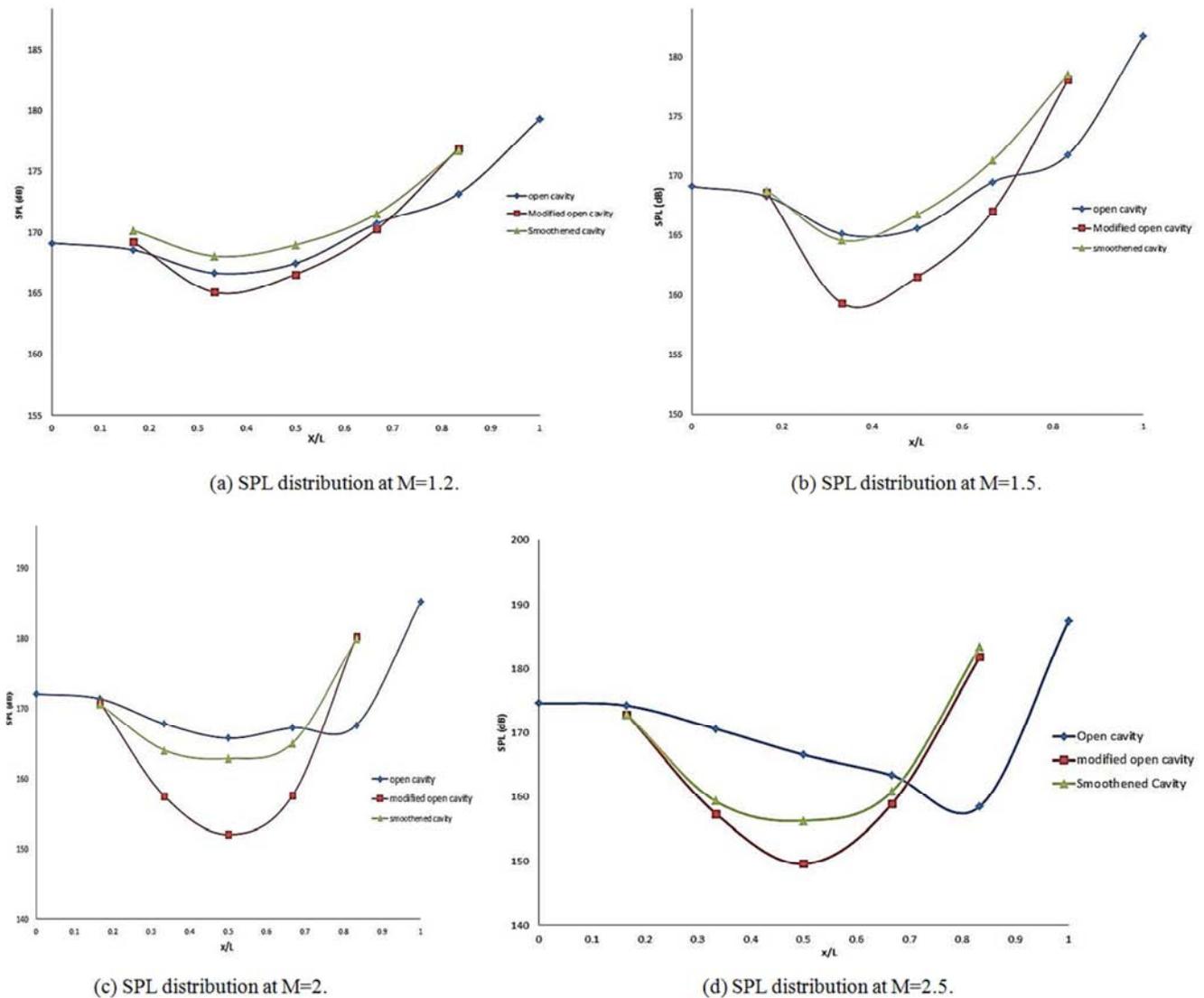

*Figure 11.* Comparison of SPL distribution across the open cavity and Modified open cavity at different Mach number.

Now, for the unmodified cavity, we see that as we increase the Mach number, the generation of acoustic noise is also increased giving rise to a high value of SPL. So, another motivation behind the modification of the initial cavity was the decrease in the value of the SPL.

After the modification, the first thing to be noticed is that for the modified cavity the value drops drastically inside the cavity in comparison to the unmodified one. Added to that, the minimum value of SPL for the modified cavity drops further with the increase in Mach number and also the position where this minima occurs shifts right with the increase in the Mach number. So sharp slant edges not only causes an overall decrement of SPL, it also causes this decrement to increase with the flow Mach number. (for example at Mach 1.5, the least value of SPL is 158 at x=0.3 and at Mach number 2.5 the minima has a value of 150 at x=0.5).

However we see that when we smoothen the corners, the SPL value has increased with respect to the sharp edged modified cavity. In fact for Mach number 1.2, SPL has values slightly more than that of the initial rectangular cavity. However this value decreases as Mach number is increased. For Mach number 2 and 2.5 the SPL values are much lower than that of the initial rectangular cavity although it never quite reaches the low peaks reached in case of the sharp edged modified cavity.

# 6. Conclusion

A numerical investigation has been conducted on open rectangular cavity and then on modified open cavities for supersonic flows with Mach number varying from 1.2 to 2.5. A comparison between the pressure contours, streamline pattern, $C_P$ distribution and SPL had been conducted between the cavities for different Mach numbers. The conclusions that can be drawn are:

The central strong vortex is stretched towards the trailing face with increment in Mach number for the open unmodified cavity but for the modified ones this vortex has a central position throughout.

The $C_P$ has a positive value for all Mach numbers for the unmodified cavity. But after modification, for both the sharp and the smoothened cavities, this value decreases drastically. The decrement keeps increasing as Mach number increases up to Mach number 2. The decrement is more for the smoothened cavity than for the sharp edged one. It attains a negative value at Mach number 2.5 thus letting us hypothesize the presence of a critical Mach number resulting in the change in direction of the circulation.

The decrease in $C_P$ indicates a decrease in drag generation inside the cavity. Thus this cavity modification (slant edges and smoothened corners) can be used in case of supersonic aircrafts for the cavities open to supersonic flows over them.

The unmodified cavity has increasing SPL with increasing Mach number. But on introducing the slant edges, a steady decrease in SPL is noticed with increasing Mach number. However on smoothening the edges the SPL has increased in value although the decrease with increasing Mach number continues. Thus a compromise has to be reached when it comes to achieving decrement in drag and as well as SPL levels.

This study can also be used to further analyze and understand supersonic combustion and subsequent flame holding techniques. Also, as we have considered a turbulent flow, we can apply this study to investigate turbulent flame and its sustainability in a turbulent environment.